\documentclass[superscriptaddress, amsmath,amssymb, aps, twocolumn ]{revtex4-1}

\usepackage{graphicx}
\usepackage{dcolumn}
\usepackage{bm}
\usepackage{wasysym}
\usepackage{hyperref}
\usepackage[mathlines]{lineno}
\usepackage{xcolor}
\usepackage{lipsum}

\usepackage{gensymb}
\usepackage{enumitem}
\usepackage{multirow}
\usepackage{booktabs}
\usepackage{array}
\newcolumntype{P}[1]{>{\centering\arraybackslash}p{#1}}
\usepackage{amssymb}
\usepackage{adjustbox}
\usepackage{orcidlink}
\usepackage{float}

\begin{document}

\title{The Era of Precision in Computational Models of Gravitational Waves}.

\author{Ulrich Sperhake 
\orcidlink{0000-0002-3134-7088}}
\email{U.Sperhake@damtp.cam.ac.uk}
\affiliation{DAMTP, Centre for Mathematical Sciences, University of Cambridge, Wilberforce Road, Cambridge CB3 0WA, UK}
\affiliation{{TAPIR 350-17, Caltech, 1200 E. California Boulevard,
Pasadena, California 91125, USA}}
\affiliation{Department of Physics and Astronomy, Johns Hopkins University, 3400 North Charles Street, Baltimore, Maryland 21218, USA}

\date{\today}

\begin{abstract}
Einstein's equations of general relativity are one of the most
complicated set of equations in all of physics and, for all
but idealized physical settings, can only be solved by numerical
methods on high-performance computing systems. Generating
such solutions is a veritable Odyssey in its own right
with adventures across the fields of mathematical
theory, physical interpretation and computing challenges.
These endeavors came to fruition in the mid 2000s when the
two-body problem of general relativity was finally solved.
And not too soon, as these results and their
follow-up investigations came to play a key role in the
Nobel-Prize winning discovery of gravitational waves by
LIGO in 2015.
\end{abstract}

\maketitle

\section{Introduction}

When Albert Einstein published his theory of {\it general relativity} (GR)
in 1915 \cite{Einstein:1916vd}, he did not merely present a refined and more accurate
theoretical description of gravitational phenomena. He certainly did
that, too, and was able to resolve astrophysical puzzles such as
the perihelion precession of Mercury, the innermost planet of the
solar system. But his theory can be regarded as nothing short of
revolutionary in how it changed our viewpoint of the very fabric
of space and time. Physical theories, prior to relativity, operated
like a script for a theater play unfolding on a stage. The theory
is formulated in the form of a set of mathematical
equations which determine how a physical system with given
initial state evolves in time or what equilibrium state a system
assumes given some boundary conditions. The solutions to these equations
consist of {\it functions} that provide for every point in space and
time the values of relevant physical quantities. A familiar example
is the weather forecast which, taking today's weather as initial
data, computes temperature, pressure, humidity etc as a function
of space (town, village or region) and future time. Space and time
merely operate as a coordinate system here; we intuitively know
what they are. Or so we thought...

Einstein already uncovered a remarkable connection between space and time
when he published his theory of {\it special relativity} (SR) in 1905.
Based solely on the experimental observation that the speed of light
is a universal constant, Einstein concluded that
the length of time intervals or physical bodies inevitably depends
on the motion of an observer. In a sense that can be made mathematically
precise, we can trade time for space or vice versa by traveling.
Despite this remarkable connection, spacetime -- as we shall
henceforth call the union of the two -- retains a
conventional nature in SR; it still forms a static background on
which events unfold but which is not affected by the actions
of the play.

Einstein's general relativity pushed this conventional viewpoint
over the cliff once and for all. Instead of merely forming the stage,
spacetime joins the cast and becomes a dynamical actor of the play
we call physics. If we identify the static viewpoint of space and time
with a football pitch on which the 22 players execute their
business, the new viewpoint is that of a stormy ocean which massively
backreacts on (and occasionally jeopardizes) the integrity and motion
of seafaring vessels. Of course, the dynamical nature of the ocean
surface is determined by various external factors such as wind
or tidal effects from moon and sun. In general relativity, we
have no external agents. Instead, the ``ocean'' represents all
spacetime and our ``set of ships'' includes wind, moon and sun
as well as every other object
that moves in spacetime and ``stirs the ocean''. The
disturbances in the ocean are thus exclusively generated by the ships
while the ocean determines how the ships move. Or, in John Wheeler's
famous quote, {\it Spacetime tells matter how to move; matter tells
spacetime how to curve.}

The branch of mathematics that enabled Einstein to formulate this
idea inside a comprehensive physical theory is called {\it differential
geometry}. We are all familiar with elements of {\it flat} or {\it
Euclidean geometry} from school; for example we ``learned'' that
two straight lines cross either once or (when parallel) not at all,
and that the angles of triangles always add up to $180^{\circ}$.
These statements are all fine and well as long as we do our geometry
with a pencil on a flat sheet of paper. But over time,
mathematicians realized that Euclidean geometry is merely one
special case and that the geometry on curved surfaces can behave
rather differently. Probably the most conspicuous example is the
surface of planet Earth. The angles of the triangle marked by the
North Pole, Singapore and the Ecuadorian capital Quito (both cities
are almost right on the equator), add up to way more than $180^{\circ}$
(try it on a globe if in doubt). Differential geometry is
the mathematical formulation of geometry on arbitrary surfaces.
Developed in large parts by Bernhard Riemann in the 19th century,
it was well known to Einstein who realized how it facilitates
the incorporation of gravity into special relativity and, thus, spawn
the theory of general relativity.

We can conceptually understand how this works by returning to
our example of the Earth's surface. Consider for this purpose
two adventurers, one at Quito and one at Singapore, and
each equipped with a perfect compass. Let us
assume the magnetic and geographic North pole coincided exactly
(rather than only approximately), and that our adventurers can walk
through mountains and across water. Our adventurers now move
straight northwards, each guided by their perfect compass.
Initially about $20000\,{\rm km}$ apart, they will inevitably
hit each other when their paths cross at the North Pole.
This is exactly what happens when the Earth and a meteorite,
initially far apart, collide under their gravitational attraction.
But there is no attractive force acting between our adventurers;
rather they are moving closer to one another because they move
on a curved surface. A fascinating and entertaining
introduction to the intricacies of geometry in curved spaces
with plentiful graphical illustration can be found in
Jean-Pierre Petit's {\it La G{\'e}om{\'e}tricon} \cite{Petit:1980}.

Einstein's theory likewise describes gravity not as a force but
as the manifestation of objects moving on seemingly straight lines,
so called {\it geodesics}, in a curved spacetime. In Einstein's theory
we have four dimensions, three for space and one for time, but
the geometrical concept is identical to our example above.
To complete his theory, Einstein now needed one further ingredient,
a set of equations that tells us exactly how the spacetime
curvature is related to the matter content. The type of equations
encountered in physics typically do not just involve unknown
variables and numbers, but they relate functions and their
rate of change. A simple example is the time evolution of
foam in a glass of beer. Foam consists of many bubbles each
of which has the same probability of bursting. The number of
bubbles we lose in each time interval is therefore
equal to a constant (between 0 and 1, to be precise)
times the current number of bubbles. In words, we look for a
function whose rate of change in time is a constant times itself.
The solution is an exponential function, so beer foam decays
according to an exponential law. The same law
governs radioactive decay.

Unfortunately, Einstein's equations are much more complicated.
First, we have 10 equations instead of one and also 10 unknown
functions instead of the single one for the beer foam. Next,
the equations involve the rate of change in space as well as
time and even involve the rate of change of the rate of change
(second derivatives in mathematical terms). Finally, the equations
involve thousands of terms and
have products of terms all over the place which makes them
{\it nonlinear}. In mathematics, the term nonlinear evokes
about the same sentiment as a tomb in an Indiana Jones movie:
exciting but full of deadly traps.
In fact, when he completed his theory, Einstein was skeptic
anyone would ever find a physically interesting solution.
Here, for once, Einstein was wrong, however. While serving
on the German-Russian front in World War I, Karl Schwarzschild
discovered the first exact solution of the Einstein equations
\cite{Schwarzschild:1916uq},
the spacetime of a point mass in spherical symmetry. This
solution later became
better known as the Schwarzschild black hole and had (and still has)
tremendous impact on deepening our understanding of Einstein's theory.
Where did Einstein err with his pessimism? The answer consists in
one word, {\it symmetry}. The Einstein equations greatly simplify
if all functions depend on the radius only, rather than on
all four space and time coordinates. In fact, the generalization
of the Schwarzschild black hole to a spinning black hole, which
depends on two coordinates instead of one, was discovered only about 50 years
later by Roy Kerr \cite{Kerr:1963ud}. Even though many exact solutions have been
found over the years and several of these are instrumental in modern
science, they all rely on major symmetry assumptions.

Most physically relevant dynamical systems do not fall into this
category, however, and for their description, we need to resort to other
means which involve some method of approximation. There exist
three such options. In {\it perturbation theory}, we assume
that the spacetime is close to an already known solution of
high symmetry. We then regard the deviations or {\it perturbations}
from the {\it background} solution as our dynamical variables,
assume that these are small and only keep terms up to linear
order. This approach, also widely used in other areas of physics,
has generated a wealth of valuable insight for systems as diverse
as the bulk evolution of the universe or the ringing of excited black holes.
A second approximation simplifies the field equations themselves,
as for example in the {\it post-Newtonian}
formalism \cite{Blanchet:2013haa,Damour:2016gwp}. Here,
the Einstein equations are regarded as a modification of
Newtonian gravity and relativistic corrections are included
only up to some specified order in the velocity or compactness
of the objects under consideration. Finally, we may approximate
the differential equations of Einstein's theory through some
form of {\it discretization} so that solutions are represented
as data sets which we calculate {\it numerically} using
high-performance computers. Contrary to the other approximations,
this approach is applicable to every regime of general relativity
and has no accuracy restrictions besides the limits of our
computational resources.
Numerical techniques are common across many areas of
physics, as for example in computing the weather forecast,
and in Einstein's theory this branch of research is called
{\it numerical relativity} \cite{Alcubierre:2008,Baumgarte:2010,Bona:2009,
Sperhake:2014wpa}.



\section{The two-body problem and gravitational waves}
\begin{figure}
\includegraphics[width=0.48\textwidth]{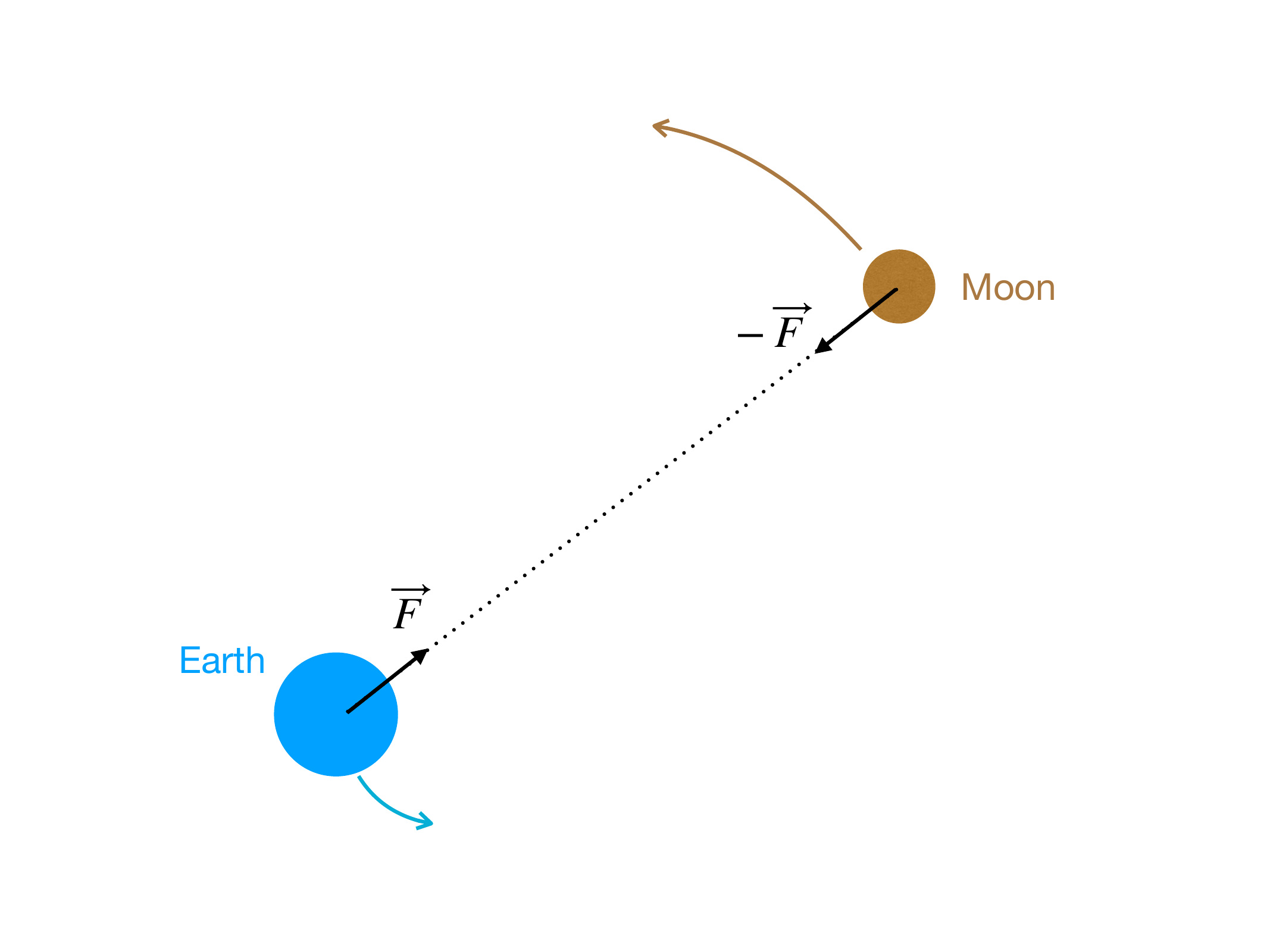}
\caption{
Illustration of the Earth-Moon system in Newtonian gravity.
Earth and Moon attract each other gravitationally with equal
force in opposite direction and orbit each other consequentially.
}
\label{fig:01}
\end{figure}

The simplest and most fundamental physical system in any theory of
gravity is the {\it two-body problem}, i.e.~the dynamical behavior of
two objects of negligible size interacting with each other exclusively
through gravity. As a classical example, we illustrate this for
the Earth-Moon system in Fig.~\ref{fig:01}.
According to Newton's law of gravity, Moon and Earth pull on each
other gravitationally with a force of equal magnitude and opposite
direction. The Newtonian equations governing this system can be solved
analytically and result in the Kepler ellipse, i.e.~the two bodies
orbit each other on a closed ellipse forever with perfect
periodicity. The same calculation can
be applied to any two objects as for example the Sun and any of its
planets. Of course, we are ignoring in such a calculation the effects
of the other planets, moons etc, and the long-term stability
of the solar system remains a topic of investigation to this
very day. The two-body problem in Newtonian physics, however,
is mathematically solved and completely understood; end of story.

In general relativity, the story is very different, and the two-body
problem was only solved in 2005 using numerical methods. To see
why this problem is much harder in GR, we need to understand
two qualitative differences between the two-body problem in GR
and that in Newtonian physics: (i) There exist no point masses in GR
and (ii) the relativistic two-body problem is {\it dissipative}, i.e.~it
gradually looses energy in the form of gravitational radiation.

Let us start with the first feature. The notion of a point particle
is naturally merely an idealization for any macroscopic object; every object
has a finite size and representing it by a point merely means that its
size is so small that any effects resulting from it can be ignored.
Mathematically, however, the concept of a point particle in Newtonian
physics is exact in the sense that we obtain a differential equation
which we can solve with perfect accuracy. In general relativity,
however, something bizarre happens that destroys the idea of a
point particle. Suppose, we have all technical skills to create
such a particle by taking a lump of matter and compressing it
into an ever smaller volume. By making the lump smaller and smaller
while keeping its total mass fixed, the curvature just outside the
surface of the lump will become larger and larger. At some point,
the curvature will become so strong that nothing can escape the region
of the lump anymore, not even light; an {\it event horizon} has formed
or, equivalently, a black hole. If we apply our experiment to the
Earth, for example, this happens when we have compressed
the entire Earth down to the size of a cherry; for the sun, we
{\it only} need to squeeze it into a sphere with radius about
$3\,{\rm km}$, roughly the size of the city of Cambridge.
This is called the Schwarzschild radius.
Even a point particle has an event horizon with Schwarzschild radius
and, therefore, a structure of
finite extent; in other words, the closest approximation to a point
particle in GR is a {\it black hole}. This has particularly dramatic
consequences when two point particles collide; gravity forces them to
merge into one single, larger black hole. In our thought experiment,
we artificially compressed matter into the event horizon, but the
singularity theorems of Roger Penrose and Stephen Hawking
\cite{Penrose:1964wq, Hawking:1967ju, Hawking:1970zqf} predict
that gravity itself is quite capable of achieving this feat without
our intervention. Even more dramatically,
their results predict that this process not
only leads to the formation of a horizon but also {\it physical singularities},
i.e.~points in spacetime where the curvature becomes infinite.

The second qualitative difference between GR and Newtonian gravity
arises from the dynamic participation of space and time in the
physical dynamics. Returning for a moment to our analogy to a
water surface, everything moving in spacetime is generating waves like
birds or fish in water. These ripples in spacetime are called
{\it gravitational waves} or {\it gravitational radiation} and their
detection by the LIGO detectors in 2015 \cite{Abbott:2016blz}
was rewarded with the
2017 Nobel Prize. The idea of gravitational waves is old and was already
noted by Einstein himself. Above we have mentioned perturbation theory as one
method of finding approximate solutions to Einstein's equations
and using this very method, Einstein realized that his equations,
when linearized, admit wave solutions \cite{Einstein:1918btx}.
Over the next 40 years,
however, the research community remained uncertain about
the physical significance of these wave solutions, and various
scientists, including Einstein, vacillated between taking
the wave solutions as real or merely a mathematical artifact.
Such confusion over the interpretation of mathematical
solutions may appear strange given the rigor of the field, but
is in fact not unusual in physics. The theory of quantum mechanics,
for example, provides mathematical results in outstanding agreement
with all types of experiment and yet Richard
Feynman famously concluded {\it I think I can safely say that nobody understands
quantum mechanics}. Readers may verify this claim by trying to understand
whether Schr{\"o}dinger's cat is alive or not.

The difficulty in understanding the reality of gravitational waves
arises from two sources. The first files under the name {\it gauge invariance}.
This seemingly complicated term means that in physics we can often
change labels or coordinates without altering anything about the
system we describe. For example, the longitude of any location of
Earth is defined relative to the Greenwich meridian. We could
switch to a new reference point, say the Estadio Azteca in Mexico
City, and all places on Earth would get new coordinates without
anything changing physically. This type of freedom appears across
all areas of physics and, given a mathematical result,
we need to carefully work out whether it merely describes such a freedom
of labeling or a genuinely observable effect. The second difficulty
in interpreting gravitational waves arises from using the
perturbation method in the first place. There is simply no
guarantee that solutions of this approximate version
of Einstein's equations actually approximate solutions of the full
equations in the weak-field limit. It often works, but we cannot be sure.

By the late 1950s the balance was finally tilting;
by combining a deeper understanding of the role of curvature in
relativity with clever thought experiments, Hermann Bondi, Richard Feynman,
Josh Goldberg,
Felix Pirani and others convincingly argued that gravitational waves
generate physically observable effects
\cite{Saulson:2010zz}. A few years later
Hermann Bondi, Rainer Sachs and co-workers settled the issue for good by
computing the effects of gravitational radiation in
Einstein's full theory using no approximations
\cite{Bondi:1962px,Sachs:1962wk}. This realization
triggered scientific investigations with major repercussions
and milestones to this day.

The first of these was the construction of devices for the direct
detection of gravitational waves. Einstein's first investigations
had already indicated that, even if real, the measurable effects
of gravitational waves passing across Earth would be tiny; in
more physical terms, the coupling of gravity to matter is very weak.
Given the potentially fatal effects of jumping off a bridge or
skyscraper building, this may appear counter intuitive at first
glance, but in a very rigorous sense, gravity is actually
millions and millions of millions times weaker than the other
fundamental interactions as for example the electromagnetic force;
a small magnet is comfortably capable of attracting a piece of metal
with a force balancing the gravitational pull from the entire planet
Earth.

In short, a gravitational-wave detector needs to be exceptionally
sensitive and accurate. More precisely, if a gravitational wave
passes through me (or any other object), it will periodically
distort my body, first higher and thinner, than shorter and thicker
and then again higher and thinner and so on. No health-risk warnings
needed, however, since the change in length is minuscule, way less
than the radius of a proton. Notwithstanding the daunting task,
Joseph Weber pioneered the construction of devices to measure
this effect by building {\it bar detectors} and even managed to
get one to the moon as part of the Apollo 17 mission. In
essence this is a metallic cylinder whose deformations can be
measured with significant precision using piezoelectric effects.
Weber eventually published claims of a positive detection
\cite{Weber:1967}, but
these results could not be reproduced and, also bearing in mind
estimates of the associated wave energy, were regarded as mere
noise artifacts. And yet, in a positive way,
Weber had opened Pandora's box.

Over the following decades,
scientists all over the world explored detector designs to
measure gravitational waves including
in particular the American {\it Laser Interferometer Gravitational-Wave
Observatory} (LIGO) \cite{LIGOweb}, its European counterparts
GEO600 \cite{GEO600web} and Virgo \cite{Virgoweb}, and the Japanese KAGRA
\cite{KAGRAweb}.
By sending laser beams back and forth
through a pair of kilometer-size
vacuum tubes in 'L' shaped arrangement, relative changes in the
length of the tubes can be measured with the astonishing
precision of about 1 part in $10^{20}$; that corresponds to
measuring the distance of the nearest star, the Alpha Centauri system
4.25 light years away, with the precision of the width of a hair.
Naturally, the construction of such a sensitive device and its
continued refinement took decades. An upscaled space version
of this concept, the {\it Laser Interferometer Space Antenna} (LISA)
\cite{LISAweb} is being build and scheduled for launch in the 2030s.

Given the tiny effect of gravitational waves, scientists quickly
realized the importance of knowing the precise shape of the
expected gravitational-wave signals. Say we have rudimentary
but far from complete understanding of some foreign language;
while we may be totally lost watching an unknown movie in that
language, we can probably listen quite effectively to the news
program since we have a fair idea what they'll be talking about.
This brings us to the second major development triggered
in the 1960s, {\it gravitational-wave source modeling}.
Already an approximate investigation
of the equations governing gravitational waves tells us that the strongest
waves are generated by physical systems with the following
attributes: (i) they must be massive and yet very compact;
(ii) they must be violent, i.e.~moving fast with high acceleration
and significant asymmetry
\cite{Peters:1963ux, Peters:1964zz}. Such objects are only found in
astrophysics and, in particular, in the form of compact binary
systems as obtained by replacing Earth and Moon in Fig.~\ref{fig:01}
with neutron stars or black holes. We have finally returned
to the two-body problem of general relativity and now
look more closely at how it can be solved.

\section{The holy grail of numerical relativity}
%
\subsection{To be or not to be well posed}

The very first seeds of numerical relativity were planted in 1952
when Yvonne Choquet-Bruhat proved that Einstein's equations are
{\it well posed} \cite{FouresBruhat:1952zz}.
To understand the groundbreaking significance of
this insight, we need to return to the concept of describing physical
systems by differential equations and, in particular, a detail
we have so far glossed over. Even if we have such a set of
differential equations, we have no guarantee that they have
unique solutions. We can obtain some understanding of the underlying
problem by considering the {\it butterfly effect} of {\it chaos theory},
whereby tiny changes in the initial conditions of a physical system
can result in
dramatic changes over time. The proverbial example is the
butterfly whose flapping of wings can alter the course of
or even generate a tornado weeks later. Importantly, the
mathematical equations governing this phenomenon or other examples
of chaos are still well posed; the time evolution sensitively depends
on the initial data and it may be hard to compute a solution in practice
but the solution is unique and fully determined by the equations.
Lack of well-posedness is a bit like chaos on steroids:
the sensitivity of the system's evolution to changes in the initial
data is not high but {\it infinite}.
This is a qualitatively worse problem than the merely practical
challenges encountered in the modeling of chaotic systems; ill-posed
differential equations do not predict the time evolution at all,
period. To make matters more complicated, the question of well-posedness
can depend on the particular way we formulate the equations or what
coordinates we choose. Yvonne Choquet-Bruhat's milestone was to
demonstrate that Einstein's equations, written in the so-called
{\it harmonic formulation}, are well-posed and
locally possess a unique solution.
This realization has been essential for numerical relativity; a given
numerical method may still fail in solving Einstein's equations
for all kinds of other reasons, but with well posedness ensured,
there should be some way to compute solutions.

With the first milestone completed, however, we straightaway hit
the next obstacle -- a pattern we will see painfully recurring as
out story unfolds. The whole idea of describing a physical system
in terms of a time evolution appears to flatly contradict the entire
spirit of relativity which Hermann Minkowski phrased as early as 1908 in
the words {\it Henceforth, space by itself and time by itself are
doomed to fade away into mere shadows, and only a kind of union of
the two will preserve an independent reality}.  This union is
commonly called {\it spacetime}, consciously written as one single
word. It is manifest in the mathematical formulation of Einstein's
equation where the distinction of space and time arises in only one
inconspicuous subtlety which we can best explain by recalling
Pythagoras' theorem $c^2=a^2+b^2$. This expression, measuring the
hypotenuse's length in a right triangle, features in
generalized form in Einstein's theory but, and this is the key
point, the square of time enters in Einstein's generalized version with a minus
sign. This appears strange as long as we merely think in terms of
triangles, but is, in fact, just a manifestation of the universal
constancy of the speed of light.  Still, the spirit of relativity
is to unify space and time, and it is a remarkable feature of the
theory that one can disentangle the two in a way that does not break
their union.  The canonical formulation of the Einstein equations
to achieve this goal was developed by Richard Arnowitt,
Stanley Deser and Charles Misner in
the early 1960s \cite{Arnowitt:1962hi}
and reformulated by Jimmy York in the 1970s \cite{York1979}; it is
commonly referred to as the ADM or 3+1 formulation of general
relativity \cite{Gourgoulhon:2007ue}
and has one further twist in the tale: Only 6 out the
10 Einstein equations are time evolution equations, the other four
impose constraints on the solution, very much like in double-entry
accounting where the sum of debits must equal the sum of credits.
If the constraints do not hold, our ``solution'' is wrong.
In relativity, this blow is softened, however: the theory guarantees
that the constraints will hold at all times provided we have
solved them for the initial data.

\subsection{The dawn of numerical relativity}
The first documented attempt to numerically compute a solution to
the two-body problem in general relativity dates back to the work
of Susan Hahn and Richard Lindquist in 1964 \cite{Hahn1964}.
They consider a spacetime
consisting of two nonrotating black holes initially at rest.  For
this configuration the initial snapshot at time zero can be described
in closed analytic form, the so-called {\it Misner data} \cite{Misner:1960zz}
found in 1960;
note that Misner's solution describes only a snapshot of
these black holes but {\it not} their time evolution.
Our physical intuition tells us (quite correctly) that the two black
holes will attract each other and eventually collide. In the 1960s,
however, the computational resources were so limited that Hahn and
Lindquist could only compute a few dozen time steps, far too short
to see more than a small fraction of this infall. Still, their work
was a pioneering proof of principle that the whole idea of numerical
relativity could work; they furthermore identified several important
challenges.

First, they highlighted the danger of numerical instability in evolving
black holes. Ultimately, computers execute only one job, albeit
with breathtaking efficiency: They compute numbers.  When we use a
computer to solve a mathematical equation, we therefore need to
convert this equation into an algorithm, i.e.~a process by which a
solution is obtained exclusively through crunching numbers.
For this purpose we need to represent smooth and continuous physical
processes in terms of finite sets of numbers.
A simple example of such a {\it discretization}
arises when we measure our age in years and ignore
hours or even finer time intervals.
We meet on our birthday
with friends or family to celebrate
the completion of yet another orbit of planet Earth around the sun.
Now suppose
the person of interest was born at 11pm on August 15 and, four
decades later stages their $40^{\rm th}$ birthday party.  Formally,
the party should not start before 11pm. In practice, we usually
ignore hours and discretize at the level of days. Measuring our
age in years or days is clearly only an approximation
to the continuous flow of time and we naturally
incur some error or uncertainty in the process; of course, we can
reduce this uncertainty by including in the description of our age
hours, minutes, seconds etc.

Computers are
much more precise in discretizing, but no matter how accurately it
is done, there remains a small error.  This brings about a fundamental
difficulty in numerical algorithms to solve differential equations:
despite looking plausible, many algorithms have the unfortunate
effect of rapidly magnifying these tiny errors, similar
to the butterfly effect we mentioned above.
This problem is by no means special to Einstein's relativity
and had already been realized by John Crank, Phyllis Nicolson,
John von Neumann and colleagues in the late 1940s and early 1950s
\cite{Crank:1947,Charney:1950},
and methods to verify the stability of simple algorithms were developed.
Due to their high complexity, however, the Einstein equations
are beyond the reach of these tests.
Furthermore, the strong curvature around black
holes affects causality in its vicinity and, thus,
introduces new ``opportunities'' for numerical algorithms
to become unstable, even if they work in more benign settings.
In the end, mathematical studies give us guidance for developing
promising algorithms, but the ultimate verification can only
be done empirically by implementing and running them in practice.

Hahn and Lindquist also noted the importance of choosing suitable
coordinates. Consider for this purpose the use of latitude and
longitude on the Earth's surface. They work fine almost everywhere,
but at the North and South pole, we cannot determine the
longitude since all meridians meet at the poles. In mathematics
this is known as a {\it coordinate singularity}, a quite separate
phenomenon from the physical singularities we encountered above. While this is
just a minor inconvenience on the surface of the Earth,
it becomes a significant problem in describing black-hole
spacetimes, especially regarding their event horizons and the region
inside.  Another challenge, also related to the coordinates, lies
in interpreting the results of a numerical relativity simulation.
Hahn and Lindquist found that the coordinate distance between their
black holes increased, contrary to our physical expectation
that they should attract each other. Coordinates, however, do not
represent real distance any better than house numbers in a street
and, simulations performed later in the century demonstrated that
the black holes indeed approach each other and collide. Clearly,
care needs to be taken in interpreting numerical results and,
where possible, coordinate invariant diagnostics are to be preferred.

\subsection{A cold case revisited}
For about 10 years, the study by Hahn and Lindquist remained the only
publication on numerical simulations of black-hole binaries.
Then in the 1970s, Bryce de Witt
initiated a reinvestigation of the problem by assigning PhD projects
to Andrej {\v C}ade{\v z}, Larry Smarr and Kenneth Eppley
\cite{Cadez:1971,Smarr1975,Eppley:1975}. They explored
the same black-hole binary system
as Hahn and Lindquist but, thanks to the enormous progress in computer
technology, with about
300 times more floating-point operations per second. Like Hahn and
Lindquist, they increased
the efficiency of their calculations by specializing to simulations
with rotational symmetry around the line of sight
connecting the two black holes -- this is called {\it axisymmetry} --
and developing an elaborate coordinate
system adapted to the shape of the black holes. Despite all efforts,
however, their evolutions remained plagued by numerical instabilities
which terminated the simulations before significant insight into
the dynamics of the black-hole binary could be obtained.

In the late 1970s,
Smarr and York \cite{Smarr:1977uf}
returned to the simpler scenario of simulating a
single black hole. The numerical modeling of a physical system we
already understand at analytic level may not lead to new physical
insight, but it provides an ideal setting to explore the deeper
reasons of the seeming failure of the then state-of-the-art numerical
techniques. Smarr and York's 1978 paper indeed provided the first
major step forward. Previous work had developed quite sophisticated
coordinates for the initial data, but did not
consider in equal detail how the coordinates evolve during the
simulation. To understand this better, imagine for the moment that
we actually live in the spacetime under consideration
and put a spacecraft at every point.
Instead of computing the future evolution with a computer, we just let
the spacecrafts trace out their world lines and measure at each
moment of their journey the geometry in their neighborhood. At the
end, we collect the data from all spacecrafts and know the entire
spacetime. We can even control to some extend how exactly our
congruence of spacecrafts moves through the spacetime, say some a bit
more to the left and others to the right, or some a bit slower than
others. This speed control represents our freedom to
choose coordinates; all we require is that the spacecrafts' paths never collide.
Quite remarkably, this analogy can be put on rigorous mathematical
ground in Einstein's theory. Now suppose we have a spacetime
with a black hole.
Clearly, we do not want any of the spacecrafts to
fly into the black-hole singularity, lest the astronauts on board get
{\it spaghettified} by tidal effects.
In a numerical simulation, we pursue exactly the same goal;
hitting the singularity causes no physical harm in a computer simulation
but the computer will cease to compute anything useful once
this has happened; our code has crashed.

Smarr and York applied our spacecraft analogy with complete mathematical
rigor to a spacetime containing a Schwarzschild black hole and showed
that for the simple time coordinate used in previous simulations,
it takes just $3.141$ (yes, the exact value is indeed $\pi$)
time units for the first
spacecraft to fall into the black-hole singularity. The
other spacecrafts, depending on their initial position, survive longer
but eventually they all suffer the same fate. Further to this insight,
Smarr and York also demonstrated how one can avoid this disastrous
outcome by switching to a better time coordinate. With their
recipe, all our spacecrafts use their rockets to accelerate in a
way to avoid the singularity. In fact, one avoids the singularity
not so much by moving around it but by moving in a way such
that the crew ages less rapidly;
this effect is similar to the slowed passage of time near black holes
which readers may recall from the {\it Interstellar} movie.
In numerical
relativity, this ``smart'' construction of a time coordinate
is called {\it singularity avoiding slicing} and it is used,
with some modifications, with great success
to this day.
At the time, it facilitated the first (albeit short) simulation of a BH
head-collision \cite{Smarr1979} predicting that $0.1\,\%$ of the total mass are
radiated in GWs, a result that turned out to be correct within a
factor of about 2.

\subsection{The Grand Challenge}
The 1980s, while widely regarded a highly fertile decade in popular
and experimental music, saw little if any progress in numerical
simulations of black-hole spacetimes. This dearth, however, was
compensated by major progress in the calculation of initial data.
Above, we have encountered Misner's data describing in analytic
form the initial snapshot of a spacetime containing two nonrotating
black holes at rest.  In 1963, Dieter Brill and Richard Lindquist
\cite{Brill:1963yv}
found a similar data set in even slightly simpler form.  Misner
data and Brill-Lindquist data were both restricted to nonrotating
black holes at rest but in 1980, Jeffrey Bowen and Jimmy York
developed a remarkable generalization to black-hole systems with
spin and initial velocity.  Let us recall for this purpose that
Einstein's equations contain four constraints which must be satisfied
by the initial data. For black holes with zero spin and velocity,
three of these constraints are trivially satisfied since all terms
appearing in the equations vanish identically.  Furthermore, the
fourth constraint simplifies to a {\it Laplace} equation which had
been well known in mathematics for centuries and could be solved
analytically. By building on an ingenious reformulation of the
constraint equations using a so-called {\it conformal decomposition}
pioneered by Andrej Lichnerowicz in the 1940s
\cite{Lichnerowicz1944} and extended by York
in the 1970s \cite{York:1971hw}, Bowen and York \cite{Bowen:1980yu}
managed to find analytic solutions to
three of the constraint equations. Even better, their solution
contains 9 parameters for each black hole, 3 each for the hole's
position, velocity and angular momentum. The fourth constraint does
not admit analytic solutions in this general case, but it is amenable
to various kinds of numerical treatment including, in particular,
the so-called {\it puncture} solution found by Steven Brandt and
Bernd Br{\"u}gmann in 1997 \cite{Brandt:1997tf}.
Bowen-York and puncture data have
undergone some variations over time but still form the backbone for
starting many contemporary black-hole simulations.

In 1993, the endeavor to simulate black-hole binaries was
revitalized with a large-scale effort in the United States,
{\it The Binary Black-Hole Grand Challenge Project}
\cite{Choptuik:1997}.
The Grand Challenge, as it is often abbreviated, involved
over 40 researchers from 10 institutions across the entire US
with the concrete target of simulating the inspiral and merger
of black-hole binaries and compute the gravitational-wave
signals generated in the event. This goal,
often referred to as {\it The Holy Grail of Numerical Relativity},
was directly motivated by the program for detecting gravitational
waves that was clearly gathering pace around the same time.
Starting around the time, many scientists also, half tongue in cheek
and half seriously, started making informal bets on which
endeavor would be accomplished first, the numerical solution
of the two-body problem or the detection of gravitational waves.

The Grand Challenge project, endowed with vastly superior
resources compared to the investigations of the 1960s and 1970s,
achieved progress in numerous aspects of numerical relativity.
By incorporating the insights gained in previous work, it
developed a new code infrastructure and simulated the first
complete collision of nonrotating black holes starting at
larger separation from rest \cite{Anninos:1994gp}.
The resulting gravitational wave signal consists of a burst
generated at merger followed by oscillations with rapidly
decreasing amplitude, a {\it ringdown} in complete analogy
to the sound a bell makes when we hit it with a hammer.
The codes were also able to compute the horizons of the
black holes during the infall and merger and extended
their studies to black holes of unequal mass
\cite{Anninos:1998wt}. But their
simulations only considered nonrotating black holes
starting from rest which means they fall towards each other in
a straight line, undergoing a {\it head-on collision}.
While helpful for exploring basic dynamics in black-hole
mergers, this scenario is a far cry from the quasi-circular
inspiral of orbiting black holes expected in astrophysics.
The holy grail of an orbiting black-hole binary is much
more challenging for a variety or reasons.

First, orbiting black holes cannot be simulated in the axisymmetric
setting used in all codes up to that point but require simulations
in fully three-dimensional space. For this purpose, the
Grand Challenge developed the first numerical relativity
code in full 3+1 dimensions named the
{\it G--code} \cite{Anninos:1995am}. It was successfully calibrated
by reproducing its predecessor's results on head-on collisions.

The second major challenge arising in the modeling of
black-hole inspiral is its long duration.
For the Grand Challenge this turned out a much bigger
problem than just needing more computers. While the
short simulations of black holes colliding head on could
just be completed, any attempts of evolving black holes longer,
either in binaries or single ones, led to
stability issues similar to those
encountered in the work of the 1960s and 1970s.
All attempts at refining the coordinate choices or improving
the numerical algorithms failed so comprehensively
that the conclusion was inescapable: Something more
fundamental was wrong.

\subsection{New horizons in a New Millennium}
Despite falling short of its main goal to reach the holy grail, the
Grand Challenge paved the way for future developments through various
key achievements. First, it brought plentiful new insights into
what works and, perhaps more importantly, what did not work.  Second,
it led to the training of a much larger community of researchers
than the field had ever seen before. Finally, the Grand Challenge
sparked the idea of {\it Community Toolkits}, i.e.~publicly available
coding infrastructures that provide key elements of modern computing
like parallelization, data I/O and discretization procedures. Rather
than writing their own code from scratch, researchers all over the
world thus became able to add their own specific numerical relativity
techniques to an otherwise ready-to-use infrastructure. The first
software package of this kind was the {\it Cactus Computational
Toolkit} \cite{Goodale2002} originally developed by Paul Walker and
Joan Mass{\'o} from 1997 on, but rapidly gathering many more
contributors, especially from Ed Seidel's numerical relativity group
at the Albert-Einstein Institute in Potsdam.  Rebranded the {\it
Einstein Toolkit} \cite{EinsteinToolkit}, it is heavily used and
developed to this day, and also sparked the development of alternative
packages.

The comparatively centralized era of the Grand Challenge was followed
by a more free-for-all period as the numerous smaller
groups and individual researchers, equipped with new
toolkits, started reaching out in all directions to explore their
renegade ideas about what fundamental problem was inhibiting
further progress in black-hole modeling. And this leads us back
to our discussion of the well-posedness of differential equations.
As proven by Choquet-Bruhat, the Einstein equations are well posed
when written in the harmonic formulation, but this is not necessarily
the case when we write them in a different formulation. I am not aware
of a clear analogy of this particular idiosyncrasy of differential equations,
but we can see through a simple example how easily an unfortunate
rewriting of an equation can lead to trouble. Consider for this purpose
the trivial equation $x=2$. This is all fine, but now we square both
sides and rewrite the equation as $x^2=4$. Even though we have performed
what looks like a harmless operation, we have obtained
a different equation that
allows for a second solution, $x=-2$, which did not exist in the
original version. The subtleties that distinguish well-posed
and ill-posed formulations of Einstein's equations are a good
deal more complicated and less evident, but the upshot remains,
given a new formulation, we have no guarantee it is well posed.

Researchers were quite aware of the fact and investigations
into the well-posedness of 3+1 formulations of the Einstein
equations started as early as the 1980s. By the late 1990s,
the ADM equations had been analyzed in various settings with
the conclusion that, in general, they are not well posed but,
in more specific mathematical terms, only {\it weakly hyperbolic}
\cite{Alcubierre:2008}.
The search for alternative, well-posed formulations of Einstein's
equations illustrates a completely separate way in which modern
computing has come to aid the modeling of physical systems,
{\it symbolic manipulation}. Exploring new formulations of
Einstein's equations, while crucial for advancing numerical
computations, is in itself a purely analytic task. Manipulating
equations as long and complex as those of Einstein's relativity
with pen on paper
is extremely laborious and fraught with dangers of making
mistakes. Today, software packages like
{\sc Mathematica} or {\sc Maple} enable us to process
enormous equations free of error.

The most successful investigations of alternative formulations
of the Einstein equations proceeded along two main directions,
one modifying the harmonic formulation employed in Choquet-Bruhat's
initial study, the second modifying the ADM equations by applying
a conformal decomposition. The harmonic
formulation employs, by construction, a specific coordinate choice,
{\it harmonic gauge}. This fixing prevents us from adapting
our coordinates to the idiosyncrasies of the specific physical
system under consideration. Attempts to generalize the approach
to generic gauge while preserving the advantages of a harmonic
formulation date back to Helmut Friedrich's work in the 1980s
\cite{Friedrich:1985afv} and,
further developed by David Garfinkle, Carsten Gundlach,
Frans Pretorius and coworkers in the early 2000s
\cite{Garfinkle:2001ni,Pretorius:2004jg,Gundlach:2005eh}, led to the
{\it generalized harmonic gauge} (GHG) formulation. Independently,
Thomas Baumgarte, Stuart Shapiro, Masaru Shibata, Takashi Nakamura,
Kenichi Oohara and Yasufumi Kojima developed a modification of the
ADM equations by applying a conformal transformation analogous
to the York-Lichnerowicz split we have encountered above in our
discussion of initial-data construction
\cite{Nakamura:1987zz, Shibata:1995we, Baumgarte:1998te}. This formulation,
commonly referred to as BSSNOK (sometimes also BSSN for short),
empirically demonstrated its superior stability properties
relative to the ADM equations in the late 1990s and early 2000s
when it was successfully employed in the numerical modeling of
single black holes and head-on collisions
\cite{Alcubierre:2000yz,Alcubierre:2002kk,Sperhake:2005uf,Fiske:2005fx}.
These simulations
worked over long times without encountering
the instabilities that had plagued their predecessors and
analytic studies eventually confirmed the well posedness of the BSSNOK
formulation. But one
problem remained: Simulations of two black holes circling around each
other, despite being pushed to completion of one orbit by
Bernd Br{\"ug}mann and colleagues in 2004 \cite{Bruegmann:2003aw},
kept crashing before their
dynamics and gravitational-wave emission
could be unraveled. The holy grail seemed to remain elusive --
but not for much longer.

\subsection{The 2005 breakthrough}
At the start of 2005, the numerical relativity community still
faced the challenge of accommodating in a numerically stable manner,
the prolonged movement of the two black holes orbiting each other
and their final merger into one spinning remnant. Moving objects
are no problem in ``normal'' numerical settings, but black holes
contain singular points in their interior and their movement is
much more difficult to handle.
Over the years, two main approaches
had been developed and used for this purpose. First, with
{\it comoving coordinates}, we can absorb the rotation \cite{Bruegmann:1997uc},
very much like we fix longitude
and latitude on the surface of the Earth, notwithstanding its
rotation. We need to take into account some nontrivial
consequences like the Coriolis force, but otherwise co-rotating coordinates
work quite well as long as we remain on the Earth. They become
quite problematic, however, when we wish to simultaneously describe
objects far away; the positions of Pluto or the Voyager spacecrafts,
for example, become a veritable mess when expressed in terms of
latitude, longitude, time and their distance from the Earth's surface.
The second approach to accommodate black-hole motion in numerical
simulations, attributed to Bill Unruh \cite{Thornburg1987}
(see also \cite{Anninos:1994dj}),
uses a procedure known as {\it excision} whereby we cut
out a region around the black-hole singularity. This can be done,
since the interior of a black hole is hidden behind a horizon anyway;
it is not visible from the outside or, more precisely, whatever
is happening inside the horizon cannot causally affect any event
outside.


\begin{figure*}[t]
\includegraphics[width=0.49\textwidth]{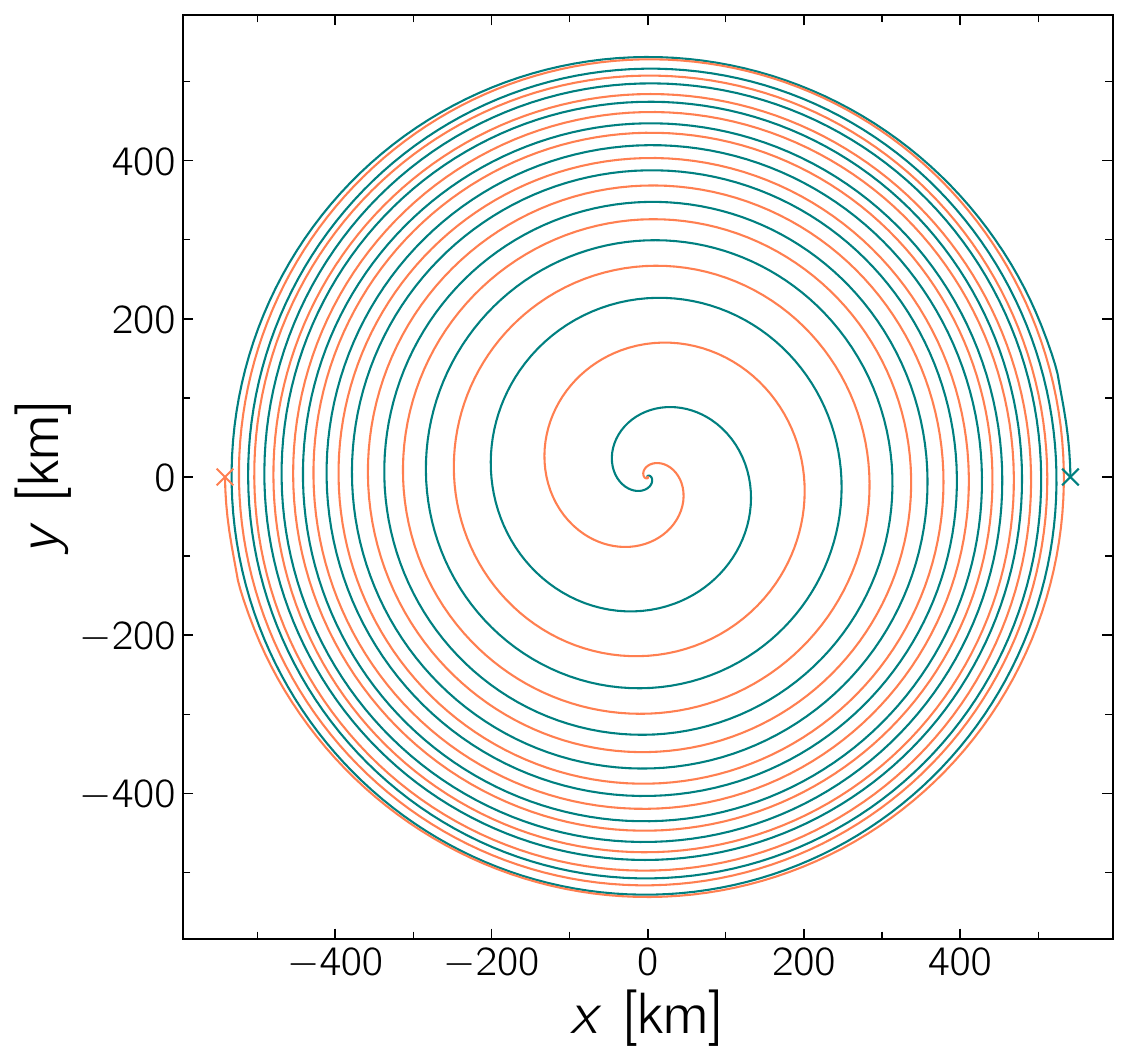}
\includegraphics[width=0.49\textwidth]{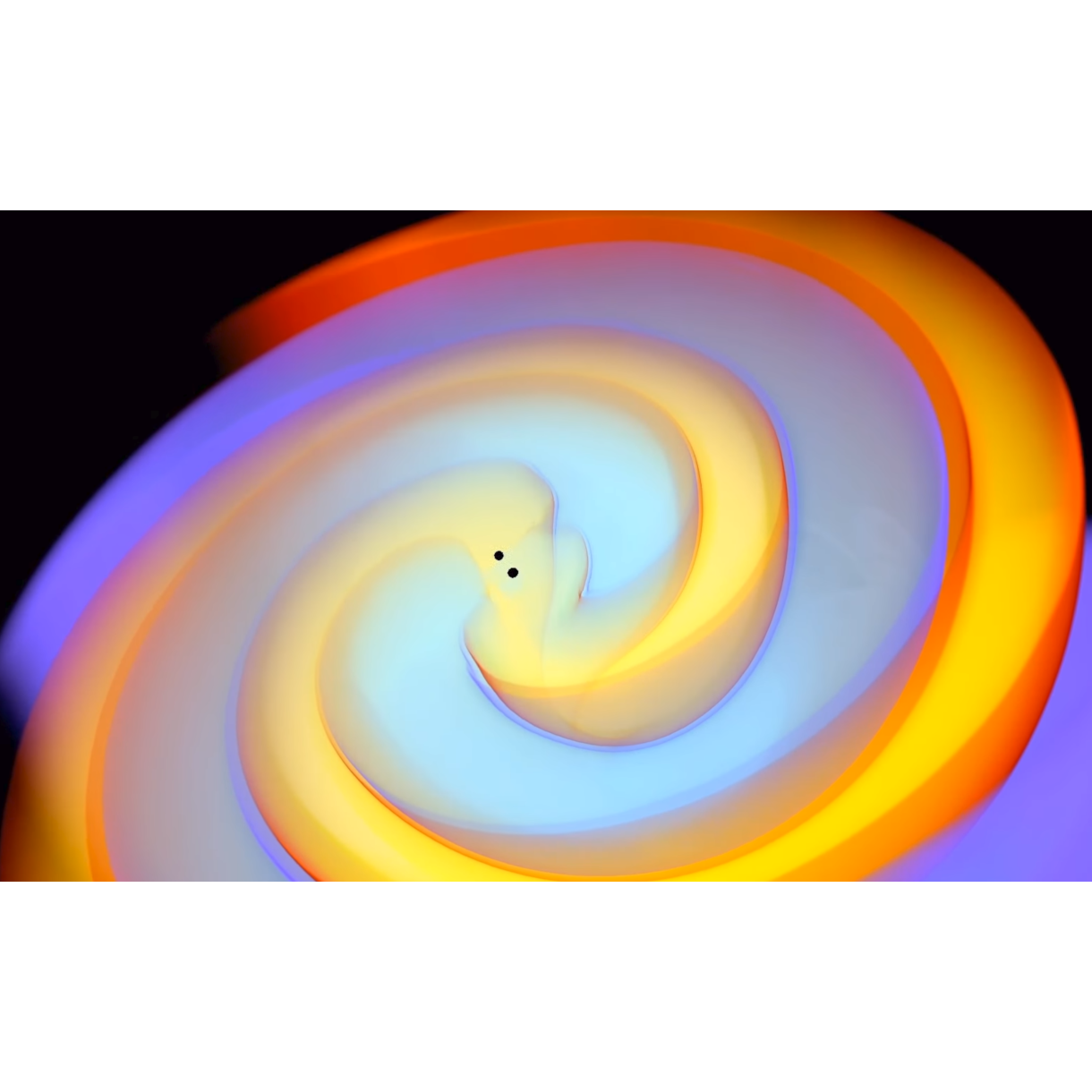}

\includegraphics[width=0.95\textwidth]{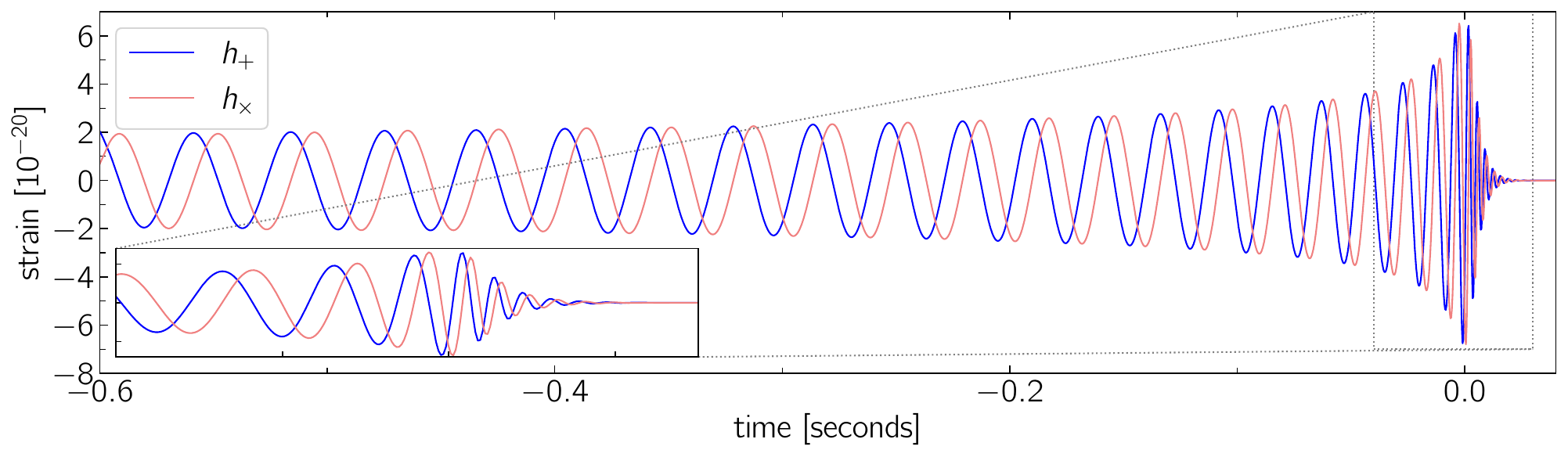}
\caption{
{\it Top left}: The trajectory of two black holes,
each 30 times as heavy as the sun,
orbiting around each other and gradually inspiraling
towards merger due to gravitational-wave emission.
The black hole on the left (orange cross) starts with velocity pointing
down, the one on the right (teal cross) with upward velocity.
{\it Top right}: A snapshot of the numerical simulation displaying
the two black holes shortly before merger together with the
gravitational-wave ripples created by their motion.
[Image credit: Markus Kunesch, Saran Tunyasuvunakool, Pau
Figueras (GRChombo); Paul Shellard, Juha Jaykka (Centre for
Theoretical Cosmology, Cambridge); Carson Brownlee, James
Jeffers, Ingo Wald (Intel Advanced Visualization and Rendering).]
{\it Bottom panel}: Amplitude of this gravitational wave as
measured on Earth if the binary were located in
the M87 galaxy of the Virgo cluster about 53.5 million
light years away. Time is measured relative to the
moment of maximal intensity. The inset shows the
ringdown when the black holes have merged into one.
}
\label{fig:02}
\end{figure*}
The breakthrough finally came in April 2005 when Frans Pretorius
presented the first multi-orbit inspiral, merger and ringdown of a
black-hole binary with a complete gravitational-wave signal
\cite{Pretorius:2005gq} at a
conference in Banff, Alberta. By combining black-hole excision with
specifically tailored coordinate conditions and adjusting the GHG
formulation according to suggestions of Gundlach and colleagues
\cite{Gundlach:2005eh},
he overcame all remaining obstacles that had hitherto kept the holy
grail out of reach. Even though a sense of progress in numerical
relativity had been in the air, the comprehensiveness of Pretorius'
success broke like thunder with an echo that can still be felt
reverberating around the community.  When we review history -- science
or other -- the developments are presented in such time compression
that one may easily overlook the tormenting effect of a decades
long arduous endeavor dedicated to reach one goal. The catharsis
of numerical relativity, however, had another twist in its tail.
About 6 months later, the numerical relativity groups of
UT Brownsville \cite{Campanelli:2005dd}
and the NASA Goddard Space Flight Center \cite{Baker:2005vv},
independent of each other, presented a second breakthrough
in two talks at another numerical relativity conference held at Goddard.
Quite astonishingly, their completion of a black-hole inspiral,
merger and ringdown, compared to Pretorius' work,
employed completely different methods in almost every regard:
They used the \mbox{BSSNOK} formalism (instead of GHG), accommodated
the black-hole motion through a different set of gauge conditions,
often referred to as {\it moving punctures}, and used other initial
data. Importantly, their results revealed excellent agreement with
Pretorius' calculations.
This not only
provides quantitative confirmation at a level of high precision, but
also demonstrated that multiple methods exist to solve Einstein's
equations, and we may choose for each problem the best suited one.
An intriguing synthesis of the beneficial properties of the GHG and
BSSNOK formulations is based on the so-called Z4 system originally
developed by Carles Bona and collaborators \cite{Bona:2003fj}, and
has been used in numerical relativity with increasing popularity
in recent years \cite{Bernuzzi:2009ex,Alic:2011gg}.

So how does the solution of the two body problem look like in
general relativity? This is illustrated in Fig.~\ref{fig:02}
which displays the motion of two nonrotating black holes,
each 30 times as massive as the sun, and the resulting gravitational-wave
strain assuming the binary is located in M87, the largest galaxy
of the Virgo cluster about $53.5$ million light years away.
We see that the black holes, starting about 1000 km apart with
initial velocities pointing up and down, respectively,
revolve around each other
in a plane similar to the Earth-Moon system. In this process,
however, the binary loses energy and their distance gradually shrinks,
the black holes {\it spiral inwards}. Eventually, they merge into a
single hole at the center. The snapshot displays the black holes
shortly before they merge together with the spacetime ripples generated
by their motion around each other. The gravitational-wave signal shown
in the bottom panel represents the fractional change in length
of our detectors' arms as a function of time when the wave finally
reaches the detector. It consists of two
contributions generating displacements in the shape of a
$+$ and a $\times$ symbol. These are shown as the blue and
orange curves; note that both have a tiny amplitude
of order $10^{-20}$. Black-hole signals of this kind exhibit
three phases: During the {\it inspiral} part
the amplitude and frequency are gradually increasing -- this part
resembles the sound of birds and is called a {\it chirp}.
The signal peaks at merger and then decreases rapidly
in the so-called ringdown.
The Nobel-Prize winning first gravitational-wave detection
GW150914 \cite{Abbott:2016blz}
was caused by a similar binary system about 20 times
further away.

\subsection{The Gold Rush Years and Gravitational-Wave Source Modeling}
Over the decades leading up to the 2005 breakthroughs, numerous
questions about black holes had accumulated in astronomy,
gravitational-wave physics and fundamental physics. Suddenly equipped
with the tools for their exploration, the numerical-relativity
community entered a {\it gold rush} era with new and sometimes unexpected
results appearing almost on a weekly basis.

Astrophysicists had been interested for a long time in a phenomenon
called {\it black-hole kicks} whereby the merger of a binary
results in preferential emission of gravitational waves in some
direction over others. The consequence is a recoil in the opposite
direction very much like that experienced in firing a gun. The phenomenon
had been known for decades, but its magnitude was unknown.
Numerical-relativity simulations revealed that the recoil velocity
could be as large as several thousand km/s for certain types of
spinning black-hole binaries
\cite{Gonzalez:2006md,Gonzalez:2007hi,Campanelli:2007ew,Campanelli:2007cga}.
This is easily large enough to eject black
holes from their host galaxies or fire them around like projectiles.

Above, we have mentioned that black holes contain spacetime
singularities, i.e.~points with infinite curvature, but that
these are hidden behind the holes' event horizons. The horizon
acts like a high-security prison, protecting the outside world
from any calamities caused by the singularity inside. Roger Penrose
conjectured that by a principle called {\it cosmic censorship}
all singularities be hidden inside a horizon \cite{Penrose:1969pc}.
Relativists had been
curious about the validity of censorship for many years. While some
exceptions had been found, these required infinite finetuning
of the initial conditions and therefore are arbitrarily unlikely
to occur in reality. Think of balancing a pencil on the tip of
your finger such that it does not fall; while theoretically possible,
you will never achieve it in practice. Interestingly,
numerical simulations demonstrate that cosmic censorship
ceases to hold once we allow for more than
3 spatial dimensions \cite{Lehner:2010pn,Andrade:2020dgc}.
In that case, two black holes merging into one
tend to produce an event horizon with fractal structure that
shrinks to zero width in finite time.

Already in 1993, Matt Choptuik \cite{Choptuik:1992jv}
discovered a feature in the formation
of black holes known as {\it critical phenomena}. This type of feature
is known from other areas of physics and bears similarity to the
phase transition of water around its freezing point. Water is liquid
at temperatures a tiny wee bit above zero but when we decrease
it marginally below zero, water qualitatively changes from liquid
to ice. The behavior of physical systems close to criticality
has some intriguing universal features common to all types of systems,
whether fluids, gravitational collapse or the behavior of compass
needles in a magnetic field. Until the 2005 breakthroughs, Choptuik's
discovery of critical collapse was regarded by some as {\it the}
result of numerical relativity. Post-breakthrough numerical-relativity
codes facilitate precision studies of a wider range of
collapse studies, including in particular scenarios beyond the
spherically symmetric case explored in Choptuik's ground breaking work.
These investigations are ongoing but already indicate a more
complex phenomenology once spherical symmetry is abandoned
\cite{Gundlach:2025yje}.

In astrophysics, black holes are expected to collide with each
other at velocities less than half the speed of light. While comfortably
enough for exorbitant speeding tickets on Earth, these velocities
are comparatively mild by Einstein's standards. Ultrarelativistic
black-hole collisions close to the speed of light, in contrast,
probe the most extreme regime of relativity and results from
numerical simulations lived up to the expectations. About $40\,\%$ of
the entire mass can be converted into gravitational radiation in this
process \cite{Sperhake:2012me} and perhaps even up to $65\,\%$
in the limit of collisions at the speed of light
\cite{Zhu:2026mhn};
for comparison, hydrogen bombs convert less than one per cent.
While not relevant in astrophysics, the hypothesized formation of
mini black holes in particle accelerators like CERN can be
theoretically studied through this type of black-hole collisions,
especially those involving extra spatial dimensions
\cite{Dimopoulos:2001hw,Eardley:2002re}. The experiments
have as yet revealed no indications that black holes can be generated
in the lab, but the numerical results on high-energy collisions of
black holes have provided valuable insight into Einstein's theory
as for example in the above tests of cosmic censorship.

The most important application of black-hole-binary modeling, however,
arises in the very field that motivated numerical relativity in the first
place: gravitational-wave modeling. To understand this, imagine
you are walking along the shore of a lake on a foggy but windless day.
You know the lake is popular with birds who are happily enjoying their
business in the middle of the lake creating all kinds of waves in the
process. But you cannot see them due to the fog. All you see are
the waves, caused by the birds' commotion,
as they reach the shore. Your task is
to infer as much information about the birds from the wave patterns
as possible, their location on the lake, their size and numbers etc.
This is very much the situation we find ourselves
in when we detect gravitational waves. To do this, we need
theoretical predictions; for each possible setup of birds --
size, location, numbers etc -- we theoretically compute how exactly
the wave signal would look like. The set of all these theoretically
possible wave signals is called a {\it waveform catalog}. We then
take the observed waveform and look for its best match in
our catalog and thus get an estimate for the most likely
bird configuration on the lake. The wave signal acts like a finger
print of the bird setup and we identify with our theoretically
computed ``finger print data base'' which one committed the act
on the lake.

In gravitational-wave astronomy, black-hole binaries are the most
common type of ``birds'' and we have already seen in Fig.~\ref{fig:02}
how the wave signal looks like if we have two non-spinning black
holes of equal mass on a nearly circular orbit. Once we allow for
rotating black holes, unequal-mass systems or eccentric orbits, we
get variations of this signal in the form of amplitude modulations,
frequency modulations and different shapes of the chirp \cite{Mroue:2013xna}.
Also, the
overall magnitude of the wave's amplitude and frequency range
strongly depend on the distance and size of the black holes. Computing
a catalog for all these possible variations is daunting task because
we have a lot of features, each quantified by one or more {\it parameters},
to consider. Even
though some of these features are comparatively easy to factor into
our modeling, a brute force approach would be prohibitively costly,
requiring millions of simulations, each using a hundred or more
cores for weeks if not longer. The task becomes much more manageable,
however, if we combine numerical simulations with the analytic
approximation methods we briefly encountered at the beginning of
our journey.

When we separated the gravitational-wave signal of Fig.~\ref{fig:02}
into three parts, inspiral, merger and ringdown, we were guided
the distinct physical phases of the binary's coalescence. The three
stages, however, also reflect regimes suitable for different
modeling techniques. While the black holes are still far apart
in the inspiral phase, they are well described by post-Newtonian
calculations. During the late ringdown, on the other hand,
we have a post-merger black hole oscillating around its equilibrium
which is well modeled by perturbation theory. The late inspiral
-- roughly the final 10 orbits --
and merger, however, are beyond such approximations and
require full numerical simulations. For each binary, results
from the three techniques can be combined into compound
wave signals, sometimes called {\it hybrid waveforms}
\cite{Ajith:2012tt,Hinder:2013oqa}. Closer
investigation of the wave signals has furthermore revealed
systematic dependence on some of the parameters which can
be exploited to construct waveform catalogs with significantly
increased efficiency; cf.~Sec.~V in Ref.~\cite{KAGRA:2021vkt}
and references therein. Despite this simplification, numerical
simulations of black-hole systems of sufficient accuracy are
challenging and costly.

Guided in particular by this challenge, the numerical-relativity
groups of the California Institute of Technology and Cornell
University started the development of a code using so-called {\it
spectral methods}. This particular discretization technique already
had achieved exquisite precision in other areas of physics, but it
was also known to have difficulties with handling singularities
present in black-hole spacetimes. In the late 2000s, following years
of arduous labor, the Caltech-Cornell group managed to combine
black-hole excision with a special version of the GHG formulation
and elaborate coordinate conditions successfully to perform the
first simulation of a black-hole binary with spectral accuracy
\cite{Lindblom:2005qh,Boyle:2007ft,Scheel:2008rj}.
Their code has undergone continuous development and is now
operated inside the Simulating eXtreme Spacetimes (SXS)
\cite{SXSweb}
collaboration involving institutes from the US and Germany.
As of writing, the SXS collaboration provides the most accurate
and also the most numerous numerical waveforms in source modeling
for the LIGO-Virgo-KAGRA network of ground-based detectors
\cite{Scheel:2025jct}.
This by no means invalidates the importance of other codes
which retain important roles for cross validation exploring
alternative types of waveform models and, perhaps most importantly,
for exploring new sources of gravitational waves; spectral codes,
while achieving superior accuracy {\it when working} tend to be
less robust in exploring new less well understood physical settings.

In historical time, our journey through the genesis of numerical
relativity has taken many decades and, paraphrasing Winston Churchill,
we have but reached {\it the end of the beginning}. The story
will undoubtedly have a sequel to be written some day in the
future and, as in any good story, our attempts at predicting its
unfolding will hardly match future reality. Still, our account
shall not end without listing some of the most exciting puzzles
the community expects to explore with upcoming gravitational-wave
observations.

{\it Dark matter and dark energy:} The type of matter
and energy we are familiar with accounts for only about
$5\,\%$ of all the content of the universe. The rest,
in absence of more descriptive knowledge, is dubbed dark matter and
dark energy. Their gravitational signature offers unprecedented
opportunities to understand what it is made of
\cite{Baryakhtar:2022hbu}.

{\it Modified gravity:} In this account we have not even attempted
to hide our fascination for Einstein's theory. And yet, contrary
to theorems in mathematics, physical theories tend to eventually
be replaced by new, more comprehensive ones. This by no means dispenses
of the previous theory as useless; Newtonian gravity, for example,
was employed with great success for achieving the moon landing.
Nonetheless, it is expected that one day, Einstein's relativity
will be replaced by a theory also encompassing quantum mechanics.
Observing the dynamics of gravity in its most extremal regime
will likely be essential for guiding these searches for
a quantum theory of gravity \cite{Berti:2015itd}.

{\it Cosmology and cosmography:} Cosmological models of the universe
rely heavily on measuring distances to far away galaxies and quasars.
Gravitational-wave observations provide a completely new way
to obtain such measures and may contribute to resolving open
questions about the evolution and origin of the universe
\cite{Aurrekoetxea:2024ypv}.
\begin{figure}
\includegraphics[width=0.48\textwidth]{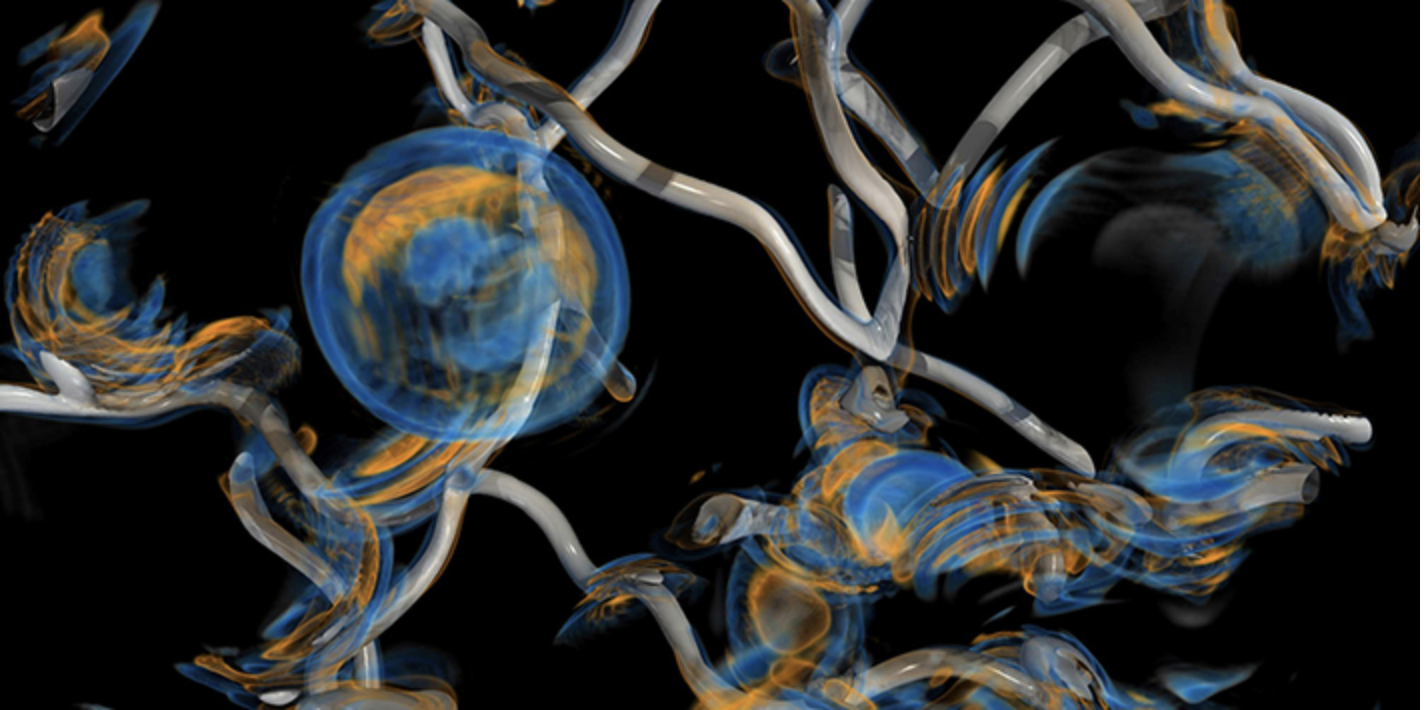}
\caption{
Snapshot of a cosmic-string network. [Image credit: Amelia Drew
(CTC, University of Cambridge) and Carson Brownlee (Intel Advanced
Visualisation and Rendering).]
}
\label{fig:03}
\end{figure}

{\it New gravitational-wave sources:} All gravitational-wave events
detected to date are compatible with predictions for black-hole
and neutron-star binaries. This interpretation, however, rests to
some extent on {\it Occam's razor} whereby, faced with multiple
explanations for on observation, the simplest is most likely correct.
Numerical relativity simulations have demonstrated that {\it exotic
compact objects}, other than black holes or neutron stars, can produce
barely distinguishable signals compared to that of Fig.~\ref{fig:02}
\cite{Evstafyeva:2024qvp}.
Furthermore, a wide range of other and quite different sources of
gravitational waves are hypothesized
\cite{CalderonBustillo:2020fyi,Clesse:2020ghq,Aurrekoetxea:2023vtp},
such as primordial
black holes, cosmic strings or domain walls. As an example, we show
in Fig.~\ref{fig:03} the snapshot of a numerical simulation of a
cosmic-string network conjectured to permeate the universe in its
very early stages. Finally, cosmological modeling increasingly employs
full numerical relativity across all stages of the Universe's
evolution \cite{Aurrekoetxea:2024ypv}.

These are merely some examples and if the history of science has
taught us any lesson, then it is to expect the unexpected. Einstein's
theory has shown us that everything in the universe jointly contributes to
spacetime, adding a new spin to the Zen inspired title
of Tony Scott's music piece {\it Is not all one?} Future gravitational-wave
explorations will undoubtedly reveal new fascinating insights into the
world we inhabit, and this will be told in a sequel to our story -- some day...

\begin{acknowledgments}
Acknowledgments: I am grateful for countless discussions, joined
projects and investigations to all the colleagues, students and
mentors I have worked with over the past decades. This article would
not exist without their inspiration. I refrain from listing everyone
individually as it would double the length of this document. Special
thanks go to Daniela Cors and Amelia Drew for proof Reading the
manuscript. This work was supported by STFC Research Grant No.
ST/V005669/1, NSF Grant No. PHY-090003, and DiRAC Projects ACTP284
and ACTP238 funded by BEIS, UKRI and STFC capital funding and STFC
operations grants.
\end{acknowledgments}
\bibliographystyle{unsrt}
\bibliography{refs_hpcgw}

\end{document}